\documentclass[%
 amsmath,amssymb,
 aps,
]{revtex4-2}

\usepackage{graphicx}% Include figure files
\usepackage{dcolumn}% Align table columns on decimal point
\usepackage{bm}% bold math
\usepackage{adjustbox}
\usepackage{multirow}
\usepackage{xcolor}

\usepackage{amssymb, url}
\usepackage[linesnumbered,ruled,lined,boxed]{algorithm2e}
\usepackage[noend]{algpseudocode}

\newcommand{\atcp}[1]{\tcp*[r]{\parbox[t]{.475\linewidth}{#1\hfill}}}
\SetCommentSty{mycommfont}

\begin{document}

\preprint{APS/123-QED}

\title{A class of randomized Subset Selection Methods for large complex networks}% Force line breaks with \\

\author{Amit Reza}
\email{amit.reza@iitgn.ac.in}
 \affiliation{Indian Institute of Technology Gandhinagar, Gujarat, India}%Lines break automatically or can be forced with \\
 %\email{richa.tripathi@iitgn.ac.in}
\author{Richa Tripathi}%
 \email{richa.tripathi@iitgn.ac.in}
\affiliation{%
 Indian Institute of Technology Gandhinagar, Gujarat, India}%

\date{\today}% It is always \today, today,

\begin{abstract}
Most of the real world complex networks such as the Internet, World Wide Web and collaboration networks are huge; and to infer their structure and dynamics one requires handling large connectivity (adjacency) matrices. Also, to find out the spectra of these networks, one needs to perform the EigenValue Decomposition(or Singular Value Decomposition for bipartite networks) of these large adjacency matrices or their Laplacian matrices. In the present work, we proposed randomized versions of the existing heuristics to infer the norm and the spectrum of the adjacency matrices. In an earlier work \cite{tripathi2019subset}, we used Subset Selection (SS) procedure to obtain the critical network structure which is smaller in size and retains the properties of original networks in terms of its Principal Singular Vector and eigenvalue spectra. We now present a few randomized versions of SS (RSS) and their time and space complexity calculation on various benchmark and real-world networks. We find that the RSS based on using QR decomposition instead of SVD in deterministic SS is the fastest. We evaluate the correctness and the performance speed after running these randomized SS heuristics on test networks and comparing the results with deterministic counterpart reported earlier. We find the proposed methods can be used effectively in large and sparse networks; they can be extended to analyse important network structure in dynamically evolving networks owing to their reduced time complexity.
\end{abstract}
\maketitle

\section{Introduction}
Most of the real world complex systems can be modelled and studied as complex networks. Also, the spectra of these networks have a direct correlation with topological properties of the networks and hence the processes on them. For example, the Internet network \cite{albert1999internet} has power law distribution of eigenvalues with tail at larger eigenvalues. Its robustness to the random failure of nodes and absence of epidemic threshold is attributed to its scale-free topology. As a consequence, the eigenvalue spectra can be used as the fingerprint of the complex networks. In many spectacular works in complex networks theory, the spectra of complex networks \cite{dorogovtsev2003spectra}, \cite{rodgers2005eigenvalue} adjacency matrix (A) and the Laplacian (L), has been successfully employed to infer synchronizability of complex networks \cite{estrada2008communicability}, \cite{sole2013spectral}, the partition of network into modules or clusters \cite{wang2008vector}, and controllability of epidemic spreading on the networks \cite{goltsev2012localization}. Similarly, the eigenvectors have been used to identify the most influential network nodes \cite{chen2012identifying}, \cite{kitsak2010identification} and spectral partitioning of networks into communities \cite{newman2006modularity}, \cite{white2005spectral}, \cite{newman2013spectral}.\\

Any algorithm that uses network topology runs on polynomial time proportional to network size. To counter this, numerous efforts have been made to represent large complex networks in a reduced form such that their spectra are preserved. This reducibility is possible because of the presence of unimportant network structure, such that the network function is robust to shutdown or failure of its nodes and links. With a similar aim, we proposed using the Subset Selection (SS) algorithm \cite{golub2012matrix} on the complex networks adjacency network \cite{tripathi2019subset} such that the obtained subset can be used to infer the spectra of original networks. This method is especially important for the cases of large complex networks where space and time complexity of analysing full adjacency matrices are too expensive. Also, although we have used the SS method to infer reduced representation of complex networks, it has been successfully used in a plethora of applications ranging from solving rank-deficient least square problems \cite{golub2012matrix}, in genetics \cite{butler2005strategies}, in wireless communication \cite{wilzeck2008antenna} and other information retrieval problems. Here we propose the randomized versions of the deterministic SS procedure and use them to infer the spectra of complex networks. \\

The Randomized SS procedures that we propose have reduced time and space complexity than the deterministic counterpart. The deterministic SS, uses first $q$ singular vectors (obtained by performing Singular Value Decomposition) of the matrix $\mathcal{A}$ (adjacency matrix for complex networks) and performs QR factorization with column pivoting on the matrix formed by these vectors to obtain a permutation vector ($P$) which is then used to reorder the columns of $\mathcal{A}$ in decreasing order of preference. The first $q$ columns of the reordered $\mathcal{A}$ form the subset. The randomized versions of SS firstly project the original matrix vectors on a lower dimensional space using a random matrix \cite{bingham2001random}. Following this, the SS procedure is performed on the projected vectors matrix. The randomized version runs on approximately half the time required by deterministic SS procedure. We also present a similar algorithm based on randomisation that offers reduced space complexity. The detailed calculation of time and space complexities for all methods are presented in the paper.

We show the results by comparing the singular value spectra obtained in the deterministic and randomized SS procedures among themselves and also with the spectra of the original network, for a predefined $q$. We also present plots of Principal Singular Vector (PSV) components for the subset obtained with deterministic and randomized procedures and capture their cosine similarities with the PSV of the original adjacency matrix. The cosine similarities that we get were fairly good depicting that the PSVs of subsets derived from the randomized versions of SS have remarkable overlap with the PSV of $\mathcal{A}$. This observation justifies that randomized subsets can competently represent the original network as far as its spectra and PSV is concerned. Also, the randomized versions select the same nodes as the important or influential ones, as chosen by the deterministic SS procedure. Hence the randomized SS offers the same benefits as deterministic SS on the complex networks in lesser time. Without loss of generality, the randomized SS algorithms can be applied to any dataset in matrix form apart from complex network adjacency matrix.\\

\section{Background}
\subsection{Mathematical preliminaries}
%{\color{red}{mathematical description of SVD, QR decomposition in short. Also about the definition of Frobenius norm in terms of $\sigma_i$}}
\begin{itemize}
 \item \textbf{Rank} : For a matrix with $m$ rows and $n$ columns, rank is defined as $r = min(m, n)$. If there is collinearity in the dataset (matrix), rank is the number linearly independent rows vectors or the number linearly independent columns vectors in the matrix, whichever is smaller.  
 \item \textbf{Energy of Matrix}: Also given by the Frobenius norm, the energy of a matrix is the square root of the sum of the absolute square of all its elements.
 It is also given by square root of sum of squares of absolute singular values  $\sigma_i$'s of the matrix.
 \begin{equation}
  \parallel A_{m \times n} \parallel_F = \sqrt{\sum_{i=1}^{i=m}\sum_{j = 1}^{j=n} |a_{ij}|^2}  = \sqrt{\sum_{i = 1}^{min(m,n)} \sigma_i^2}
 \end{equation}

 \item \textbf{Singular Value Decomposition (SVD)}: The Singular Value Decomposition is a matrix factorization technique and is a generalization of eigen decompostion to non-square matrices.
 For a matrix $A_{m \times n}$, the SVD is given by $M = U \Sigma V^T$, where $U_{m \times m}$ and $V_{n \times n}$ are orthogonal matrices with the left and the right singular vectors of $A$ and $\Sigma_{m \times n}$ is diagonal matrix with singular values $\sigma_i$ along its diagonal. 
 
 \item \textbf{QR}: The QR decomposition of matrix $A$ is a matrix factorisation technique which decomposes matrix as $A = QR$, where $Q$ is an orthogonal matrix, and $R$ is an upper triangular matrix. 
 
 \item \textbf{QR-cp}: QR factorization of $A_{m \times n}$with column pivoting is given by,
 \begin{equation}
  A = Q \binom{R}{0} P^T, \ m \geq n
 \end{equation}
 where $Q$ and $R$ as before and $P$ is permutation matrix such that,\\
 
 \begin{equation}
  |r_{11}| \geq |r_{22}| \geq |r_{33}| \geq.....\geq |r_{nn}|
 \end{equation}
 
 and, for each $l$
 
 \begin{displaymath} \vert r_{ll}\vert \ge \Vert R_{l:j,j}\Vert _2 \quad \mbox{for $j = l+1, \ldots, n$.} \end{displaymath}

 \item \textbf{Random Projection}: It is a method used to reduce the dimensionality of data points lying in a Euclidean space. RP is generally used in handling 
 and manipulating large datasets or manifolds to infer their intrinsic dimension or to know the principal directions.

 \item \textbf{Johnson–Lindenstrauss lemma}: The JL lemma states that the dataset can be projected to a much lower dimension than the original such that the 
 distances between the points in the datasets in the original and the projected space is preserved. It is used extensively in the problems of compressed sensing,
 dimensionality reduction and graph embedding.\\
 For any $0 < \epsilon < 1$ , a set $A$ of $m$ points in $N$ dimensional space, and a number $n > 8ln(m)/\epsilon^2$ there exists a linear map 
$ \textbf{f}: \mathbb{R}^N \rightarrow \mathbb{R}^n$,  
 \begin{equation}
 ( 1-\epsilon ) \parallel x- y \parallel ^2 \leq \parallel f ( x )-f ( y ) \parallel ^ 2 \leq ( 1 + \epsilon ) \parallel x-y \parallel ^ 2 
 %{\displaystyle (1-\varepsilon )\|u-v\|^{2}\leq \|f(u)-f(v)\|^{2}\leq (1+\varepsilon )\|u-v\|^{2}} (1-\varepsilon )\|u-v\|^{2}\leq \|f(u)-f(v)\|^{2}\leq (1+\varepsilon )\|u-v\|^{2}
 \end{equation}
 for all $x , y \in A$.
 
 \item \textbf{Subset}: A subset is the part of the original matrix with only significant columns such that it has the same Frobenius norm as that of the original matrix. The subset of a matrix exists if the singular value spectrum of the dataset is such that most of the matrix norm is made up of only a few singular values. The number of these dominant singular values is governed by the matrix rank $r$.
 
 \item \textbf{Centrality}: In complex network theory, the centrality of a node or an edge is an indicator of its importance in the network. Depending on the ways
 one defines importance there are many centrality measures regularly used such as degree centrality, betweenness centrality, eigenvector centrality, closeness
 centrality, PageRank centrality, etc.
 
\end{itemize}

\subsection{Subset selection procedure}
%{\color{red}{Only need to write the steps and algorithmic form. I will add the algorithmic form.}}
%%%%=====================================================
The subset selection \cite{kanjilal1995application} procedure is a well-known method to identify the essential column vectors of a data matrix $\mathcal{A}$. In our previous work, we showed the application of the subset selection procedure in large complex networks \cite{tripathi2019subset}. The basic idea of the deterministic subset selection (SS) method is to discard the redundant columns of the data matrix and keep those columns which have a maximum contribution in the matrix in terms of energy preservation. 
Mathematically the data matrix, $\mathcal{A}$ can be thought of as a collection of two blocks $[\mathcal{A}_1, \mathcal{A}_2]$. Where $\mathcal{A}_1$ contains $q$ linearly independent columns which can approximately span the entire column space of $\mathcal{A}$ and $\mathcal{A}_2$, the collection of the redundant columns can be represented by the linear combination of the column vectors of $\mathcal{A}_1$.

It is understandable that the value of $q$ is dependent on the number of redundant columns. If the number of redundant columns is high, then $q$ will be decidedly less and vice-versa. Therefore the value of $q$ can be directly related to the numerical rank of the data matrix $\mathcal{A}$. Hence for a large data matrix, the size of the obtained subset $\mathcal{A}_1$ depends on the rank-deficiency of the data-matrix as it directly translates the value of $q$. 
To find the essential (non-redundant) block $\mathcal{A}_1$ and redundant block $\mathcal{A}_2$ respectively of $\mathcal{A}$, one has to compute the permutation matrix $P$ which helps to identify and separate out the two blocks as follows.
\begin{equation}
\label{Perm}
\mathcal{A}\, P = [\mathcal{A}_1, \mathcal{A}_2]
\end{equation}

Therefore the whole SS procedure \cite{kanjilal1995application} boils down to obtaining the optimal permutation matrix $P$. One can end up with different realisations of the $P$ matrices, but the optimal one will be decided based on the following criteria. The optimal $P$ should follow the following constraints. 
\begin{enumerate}
    \item The number of linearly independent columns ($q$)of $\mathcal{A}_1$ should represent the numerical rank of the data matrix $\mathcal{A}$.
    
    \item The residual difference between the norm of the linear combination of $\mathcal{A}_1$ with $\mathcal{A}_2$ should be minimal.
\end{enumerate}

These constraints mentioned above can be full-filled by studying the singular value decomposition(SVD) of  $\mathcal{A}$. SVD \cite{golub2012matrix} is one of the best numerical methods to obtain the critical basis of any arbitrary data-set. The singular value spectra provide the corresponding weights of the basis vectors; therefore it can be used to find out the value of $q$ as the numerical rank (normally decided based on the top-$q$ non-zero values).

Therefore the first step of the deterministic SS procedure is to obtain the factors $\mathcal{U}$, $\Sigma$, and $\mathcal{V}$ using SVD. 
\begin{equation}
\label{svd}
\mathcal{A} = \mathcal{U} \Sigma \mathcal{V}^T.
\end{equation}

The matrix $\mathcal{U}_{m \times r}$ represents eigenvectors of the left subspace of $\mathcal{A}$; the matrix $\Sigma_{r \times r}$ is a diagonal matrix with $r$ ($r = rank(\mathcal{A}) = min(m,n)$) positive non-zero entries (known as singular values) arranged in descending order of magnitude i.e ($\sigma_1 > \sigma_2 > \sigma_3 > ...> \sigma_r$) and the matrix $\mathcal{V}_{r \times n}$ represents eigenvectors of the right subspace of $\mathcal{A}$. 

To find out the value of $q$ we need to maximized the preservation of the Forbenius norm $\mathcal{A}$ such way that
\begin{equation}
\sum_{i = 1}^{q} \sigma_i^2  \simeq \sum_{i = 1}^{r} \sigma_i^2 = \| \mathcal{A} \| _F^2
\end{equation}

where $\| \mathcal{A} \|_F^2$ is the Frobenius norm of $\mathcal{A}$.\\

From eq.\ref{Perm} $\&$ \ref{svd}, it is clear that $P$ can be obtained by permuting the first $q$ columns of the $\mathcal{V}$. This implies that finding out of $q$ important columns of $\mathcal{A}$ can be equivalently translated as to a problem of the finding out the optimal permutation of the corresponding columns in $\mathcal{V}^T$. Hence the second essential step of the deterministic SS procedure is to obtain $P$ based on truncated $\mathcal{V}^T$. The complete version of the deterministic subset selection procedure is described in Algo-\ref{alg:DetSS}.

\vspace{0.25cm}
%%%%=====================================================
\begin{algorithm}[H]
\DontPrintSemicolon
\KwIn{Data Matrix \{$\mathcal{A}_{m\times m}$\}.}
\KwOut{${\mathcal{A}_1}_{m \times q}$}
$\mathcal{U}\, \Sigma \, \mathcal{V}^T = \text{svd}(\mathcal{A})$  \atcp{Compute the singular value decomposition}
$\bar{\mathcal{V}} = \mathcal{V}[:, 0:q]$ \atcp{Choose first $q$ columns of $\mathcal{V}$.}
$Q, R, P = qr(\bar{\mathcal{V}}^T)$ \atcp{Compute column-pivotal $QR$ decomposition.}
$\tilde{\mathcal{A}} = \mathcal{A}[:, P]$ \atcp{Permute the column of the data-matrix using permutation matrix $P$.}
${\mathcal{A}_1}_{m \times q} = \tilde{\mathcal{A}}[:, 0:q]$ \atcp{Choose first $q$ columns.}
\caption{\textbf{Deterministic subset selection algorithm}}
\label{alg:DetSS}
\end{algorithm}
%%%%=====================================================
\vspace{0.25cm}
It is notable that one can apply the deterministic SS procedure for a system represented by a complex network. For that purpose one has to compute the SVD of the network's adjacency matrix ($\mathcal{A}_{m \times m}$) and need to follow the steps described in Algo-\ref{alg:DetSS} to obtain ${\mathcal{A}_1}_{m \times q}$, where $\mathcal{A}_1$ is a $m \times q$ matrix representing important columns of original network adjacency matrix, $\mathcal{A}$.
%%%%=====================================================
% Suppose $\mathcal{A}_{m \times m}$ is the adjacency matrix of a undirected and unweighted network and its SVD is represented as $\mathcal{A} = \mathcal{U} \Sigma \mathcal{V}^T$, where $\mathcal{U}$ and $\mathcal{V}$ 

%  Let $\bar{V}$ represent first $q$ columns $\mathcal{V}$ and $\bar{V}^T$ is its transpose. Next, the $QR_{cp}$ factorisation of $\bar{V}^T$ is performed and a permutation vector $P$ is obtained as follows,

% \begin{equation}
% Q, R, P = QR_{cp}(\bar{V}^T)
% \end{equation}
% where $Q^TQ\ =\ I$. \\

% The $P$ vector was then employed to pick the important network sub-structure as, 

% \begin{equation}
% [\mathcal{A}_1, \mathcal{A}_2] \equiv \mathcal{A}P
% \end{equation}

% where $\mathcal{A}_1$ is a $m \times q$ matrix representing important and reduced structure of the original network matrix, $\mathcal{A}$.
%%%%=====================================================
\subsection{Bottleneck of deterministic SS procedure}

It is clear from the Algo-\ref{alg:DetSS}, two essential numerical steps are involved in the deterministic SS procedure. First is the computation of SVD of the data matrix to obtain the set of basis vectors and the second one is to apply column-pivotal QR decomposition on the truncated right singular basis to get the optimal permutation matrix. 
For a large data matrix, the computation of the SVD is a practical issue. Generally, for a data matrix $\mathcal{A}_{m \times n} : m > n$, the theoretical time complexity of SVD is $\mathcal{O}(mn^2)$. Time complexity will also increase exponentially with an increase in the size of the data matrix. Also, the first step in Algo-\ref{alg:DetSS} involves computation of the full set of right singular basis and followed by the rejection of the less important set of basis from the full set of obtained basis. Therefore, there is a wastage of computational resources. For the large data sets where the expected value of $q(\ll \text{min}(m, n))$ is minimal, the waste of computational resources will increase tremendously. For large sparse complex networks, it is quite often the scenario. Therefore, deterministic SS procedure will take a long pre-processing time to provide SVD factors. In this work, we want to investigate the possible way to reduce this pre-processing time taken by SVD by introducing the random-projection (RP) based scheme to obtain the factors quickly. In this paper, we proposed a class of randomized SS procedure which can be used to omit this bottleneck. 
The second step (line no $3$ in the Algo-\ref{alg:DetSS}) is essential to obtain $P$, therefore for both deterministic and randomized SS procedures, this step is required. 
%%%%=====================================================

\vspace{0.25cm}
\begin{figure*}[htbp]

\centering
\begin{tabular}{ccc}
 \centering
 \includegraphics[height = 3.8cm]{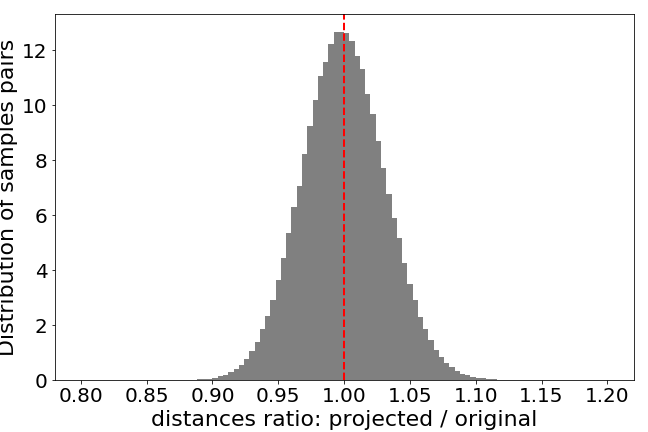}&
 \includegraphics[height = 3.8cm]{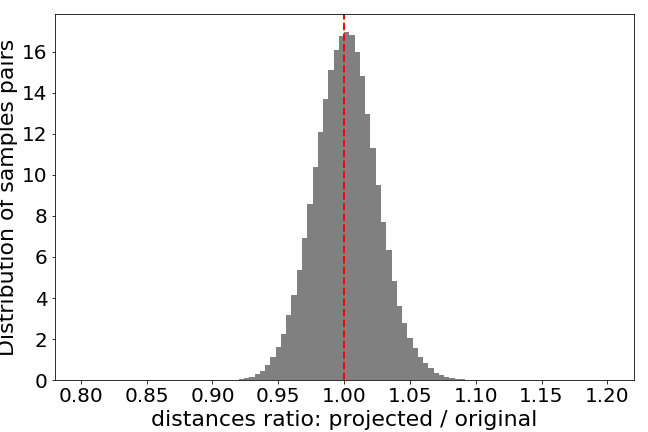}&
 \includegraphics[height = 3.8cm]{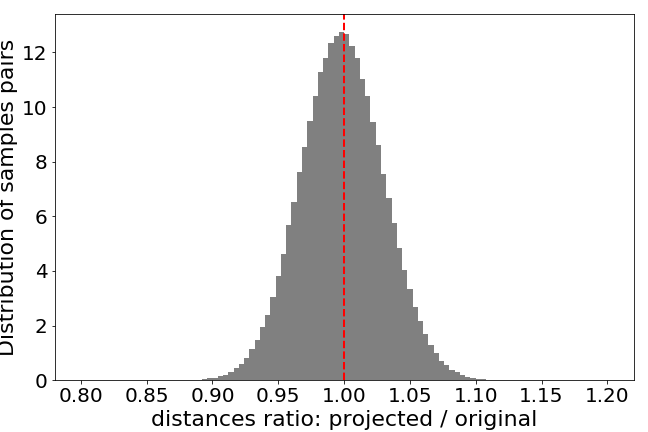}\\
 (a)&(b)&(c)
 \end{tabular}
\label{fig:RP1}
 \end{figure*}
 
\begin{figure*}[htbp]

\centering
\begin{tabular}{ccc}
 \centering
 \includegraphics[height = 3.8cm]{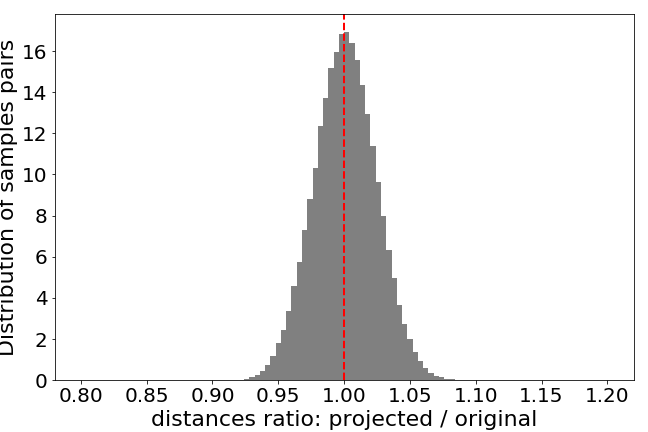}&
 \includegraphics[height = 3.8cm]{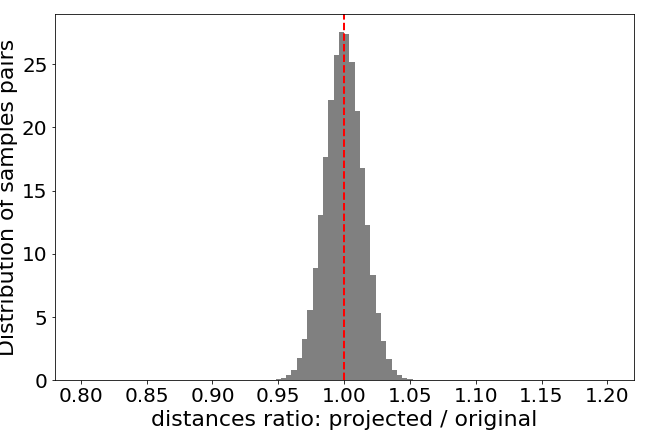}&
 \includegraphics[height = 3.8cm]{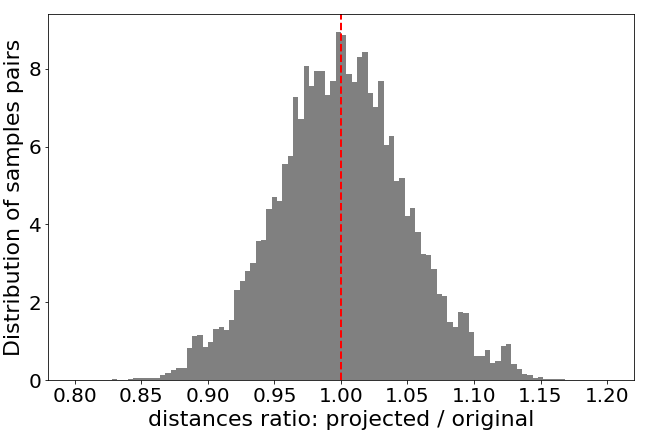}\\
 (d)&(e)&(f)
 \end{tabular}
 \caption{The histograms in figures (a)-(f) show the distribution of ratio of pair wise distances between vectors in original high dimensional space and vectors in low dimensional space. The Guassians in these plots centred around mean value 0, show that the distances are preserved in projected space as required for Random Projection method. Each of the figures correspond to $q \approx n/2$. They stand for (a) Barabasi- Albert network, (b) Drosophila network, (c) Erdos-Renyi Network, (d) Friendship network, (e) Power Grid Network and (f) US Air network respectively. }
\label{fig:RP1} 
 \end{figure*} 

\section{Randomized subset selection scheme}
%{\color{red}{Need to describe the basic algorithm and how it can help to handle large data matrix.}}
The deterministic SVD is a computationally expensive method for large data sets. Furthermore,
it is challenging to parallelize the standard SVD technique to utilise advanced computer architectures. Recently developed randomized algorithms based on RP \cite{halko2011finding} for low-rank approximation, are computationally efficient, accurate up-to a certain precision and robust. Therefore it outperforms the traditional matrix factorisation scheme in many practical problems \cite{halko2011finding}, \cite{halko2009finding}. The randomized algorithm is powerful because of the following reasons.   
\begin{itemize}
    \item The scheme is computationally efficient.
    \item The main involved operations (steps) can be optimised on modern computational architecture. 
\end{itemize}

The key concept of the randomized scheme is to exploit the randomness to construct a surrogate lower dimensional ($\bar{\mathcal{A}}_{q \times n}$) data matrix which captures the maximum information of a high-dimensional input data matrix $\mathcal{A}_{m \times n}$. The apparent assumption is that the input data matrix has a low-rank structure, i.e., the numerical rank is smaller than its original dimensions.
Without loss of generality, we assume that $n < m$, therefore to transform from original data matrix $\mathcal{A}$ to a surrogate data matrix $\bar{\mathcal{A}}$, one has to generate a random test matrix $\Omega_{q \times m}: \Omega_{ij} \in \mathcal{N}(0,1)$ and has to operate on $\mathcal{A}$ such a way that
\begin{equation}
\label{RP}
\bar{\mathcal{A}}_{q \times n} = \Omega_{q \times m} \, \mathcal{A}_{m \times n}
\end{equation}
Essentially, the operation described in Eq.\ref{RP} is known as RP of the original data vectors to a lower dimensional space. Basically, operating $\Omega$ onto $\mathcal{A}$ compressed the $n$ column-vectors ($\in \mathbb{R}^{m}$) of $\mathcal{A}$ and projected them to a lower dimension space $\mathbb{R}^{q}$. Therefore, $n$ column vectors transformed from $m$ dimensional to $q$ dimensional vectors. Since, $m$ is very large and $q$ is very small, therefore this specific transformation ($\mathcal{A} \rightarrow \bar{\mathcal{A}}$) helps to reduce the computational cost as one can directly use the transformed vectors for further use.
The point of concern is: how robust is the transformation? The RP can preserve the distance between each pair of the column vectors in the lower dimension within a $\epsilon$-error. The J-L lemma ensures this. Hence, the surrogate matrix $\bar{\mathcal{A}}$ can be used to obtain the approximated first $q$-right singular vectors ($\mathcal{V}$). Different approaches are also possible to obtain this. Algorithm-~\ref{alg:RandSS1} - ~\ref{alg:RandSS4} are the representation of a class of such randomized versions.

We show the efficacy of the RP method in case of the data of the complex network. We computed the pairwise distances of the column vectors in the original and lower dimensional space for the six different real-world and model network adjacency matrices (Drosophila, Friendship, Power-Grid, US-Air, Barabasi-Albert, Erdos-Renyl Networks). The details (number of nodes, edges) of the data are mentioned in TABLE III.
We are considering $q \approx 50 \% \text{of the nodes}$ for complex network data. Figure \ref{fig:RP1} depicts the ratio of the pairwise distances of the vectors between the lower-dimension and original feature space. If the pair-wise distances are preserved, then the ratio should be $1$. Therefore it expected that the mean of the distribution of the pair-wise distance would be $1$ and the variance will vary depending on the dimension of the projected space and number of projected vectors. Figure \ref{fig:RP1} shows that for six different complex network data, the mean of the distribution is $1$. Therefore, the projected dimension $q = m/2$ can preserve the pair-wise distance of the column vectors in lower-dimensional space with high accuracy. Hence for these data randomized SS is applicable and the retrieval of right subspace is also possible. 
One should also note that for complex network data analysis, we are applying RP onto an adjacency matrix which is a symmetric matrix; therefore the compression of the column vectors or the compression of the row vectors provides the same results. Also, although in all the algorithms we denote the input matrix as a rectangular matrix $\mathcal{A}_{m \times n}$, for the complex network data they are symmetric square matrices.  

%%%%====================================================

%%%%=====================================================
\vspace{0.25cm}

The projection in low dimension is mediated by the product of a random matrix $\Omega_{q \times m}$ with $\mathcal{A}$. The low dimension projection preserves the column norms, and the computation of any decomposition (SVD/ QR) of the matrix formed by these transformed columns gets computationally cheaper as compared to performing decomposition in original higher dimensional space. We employ this trait of RP in all proposed randomized SS procedures. In the following subsections, we present the randomized subset selection versions in details. %We also present them in algorithmic representation towards the end of the paper. {\color{blue} Write an account for left compression and right compression that we discussed}
%%%%=====================================================

%%%%=====================================================
\subsection{Randomized Subset Selection 1}
To reduce the computational complexity of the deterministic subset selection process, we propose the use of random projection to obtain top-$q$ right singular vectors. The original matrix $\mathcal{A}$ which can be conceived as $n$ column vectors in $m$ dimensional space, is projected onto a low dimensional space such that it becomes $\mathcal{\bar{A}}_{q \times n}$. In lower dimension, it transforms such that there are $n$  column-vectors in $q$ dimensional space, obtained
by a procedure as follows,
\begin{equation}
\label{Eq:svd_Y}
\mathcal{\bar{A}}_{q \times n} = \Omega _{q \times m} \mathcal{A}_{m \times n} \ ,\ k \ll m
\end{equation}
where $\Omega \in \mathcal{N}(0,1)$. 
Now one can compute SVD of $\bar{\mathcal{A}}$ to obtain the right singular vectors $\mathcal{V}$ of $\mathcal{A}$. 
Let the SVD factors of $\bar{\mathcal{A}}_{q \times m}$ are as follows
\begin{equation}
\mathcal{U}_{\bar{\mathcal{A}}} \, \Sigma_{\bar{\mathcal{A}}} \, \mathcal{V}_{\bar{\mathcal{A}}}^T \approx \bar{\mathcal{A}}
\end{equation}
The dimension of $\mathcal{V}_{\bar{\mathcal{A}}}$ is $q \times m$. 
Theoretically, one can think of 
\begin{equation}
\label{Eq:11}
\mathcal{V}_{\bar{\mathcal{A}}} = \text{eigenvector}(\bar{\mathcal{A}}^T \, \bar{\mathcal{A}})
\end{equation}

Again,
\begin{equation}
\label{Eq:2}
\bar{\mathcal{A}}^T \, \bar{\mathcal{A}} = \big(\mathcal{A}^T\Omega^T\big) \, \big(\Omega\mathcal{A}\big) \approx \mathcal{A}^T\mathcal{A} 
\end{equation}
Where we have assumed that $\mathbb{E}(\Omega^T\Omega) = \mathbb{I}_{m \times m}$ which is the inherent property of random projection matrix due to the fact the columns are almost linearly independent as they are drawn from standard Gaussian distribution.  

Using the fact from Eq.\ref{Eq:2}, one can rewrite the Eq.\ref{Eq:11} as
\begin{equation}
\label{Eq:1}
\mathcal{V}_{\bar{\mathcal{A}}} = \text{eigenvector}\big(\bar{\mathcal{A}}^T \, \bar{\mathcal{A}}\big) \approx \text{eigenvector}\big(\mathcal{A}^T \, \mathcal{A} \big)|_{\text{top}\ q \, \text{basis}}
\end{equation}
Therefore one can approximate $\mathcal{V}$ by $\mathcal{V}_{\bar{\mathcal{A}}}$ without applying SVD to the original data matrix $\mathcal{A}$. The surrogate $\bar{\mathcal{A}}$ can be used to obtain it. It is notable that only top-$q$ right singular vector of $\mathcal{V}$ can be approximated with high accuracy by $\mathcal{V}_{\bar{\mathcal{A}}}$. Other $m-q$ number of right singular vectors are not possible to retrieve. 
Hence this scheme can reduces the time required to compute the SVD as the number of columns are reduces by a factor $m-q$. The rest of the steps are similar as deterministic SS. The algorithm for the procedure is given in Algorithm-\ref{alg:RandSS1}.\\

% Then the SVD decomposition of $\mathcal{\bar{A}}$ was performed and the resulting orthogonal $V^{T}_{q \times m}$ matrix was used in place
% of $\bar{V}^T$ in the previous method {(\color{blue}section 2.0)} to find $P$ and $\mathcal{A}_1$ using $QR_{cp}$ factorization.
% \begin{equation}
% U\Sigma V^{T} = SVD (\mathcal{\bar{A}}_{q \times m})
% \label{eqn:eqsvd}
% \end{equation}

% \begin{equation}
%  \bar{V} = V_{m \times q}
% \end{equation}

% \begin{equation}
%  Q, R, P = QR(\bar{V}^{T}_{q \times m})
% \end{equation}

% The matrix subset ($\mathcal{A}_1$) is obtained by ordering columns of $\mathcal{A}$ according to the permutation vector $P$ and picking up first $k$ columns of permuted
% $\mathcal{A}$.\\
% The low dimension projection of $\mathcal{A}$ is justified as the distances between the vectors in both the spaces (original and projected) are maximally 
% preserved with a small error. {\color{blue} We show this numerically by calculating the JL bound on the distances between the vectors in original space and the distances
% between vectors in projected space [refer figure]}.
%%%%=====================================================

\begin{algorithm}[H]
\DontPrintSemicolon
\KwIn{Data Matrix \{$\mathcal{A}_{m\times n}$\}, Random Projection Matrix \{$\Omega_{q \times m } : \Omega_{ij} \in \mathcal{N}(0,1)\}$}
\KwOut{${\mathcal{A}_1}_{m \times q} $}
 $\bar{\mathcal{A}}_{q \times n} = \Omega\mathcal{A}$ \atcp{Compression of the row-space}
$\mathcal{U}_{\bar{\mathcal{A}}}\, \Sigma_{\bar{\mathcal{A}}} \, \mathcal{V}_{\bar{\mathcal{A}}}^T = \text{svd}(\mathcal{\bar{\mathcal{A}}})$  \atcp{Compute the singular value decomposition}
%$\bar{\mathcal{V}}_{\bar{\mathcal{A}}} = \mathcal{V}_{\bar{\mathcal{A}}}[:, 0:q]$ \atcp{Choose first $q$ columns of $\mathcal{V}_{\bar{\mathcal{A}}}$.}
$Q, R, P = qr(\bar{\mathcal{V}}_{\bar{\mathcal{A}}}^T)$ \atcp{Compute column-pivotal $QR$ decomposition.}
$\tilde{\mathcal{A}} = \mathcal{A}[:, P]$ \atcp{Permute the column of the data-matrix using permutation matrix $P$.}
${\mathcal{A}_1}_{m \times q} = \tilde{\mathcal{A}}[:, 0:q]$ \atcp{Choose first $q$ columns.}
\caption{\textbf{Randomized subset selection algorithm-I}}
\label{alg:RandSS1}
\end{algorithm}
%%%%====================================================
\vspace{0.25 cm}
For a system of complex network, if the network is very large, a large adjacency matrix has to be dealt with and full SVD of $\mathcal{A}$ is needed to be performed in order to find its right singular vectors $\mathcal{V}$. The above prescribed randomized SS can be useful to reduce the time complexity to obtain the SVD factors promptly.

%%%%=====================================================
\subsection{Randomized Subset Selection 2}
%{\color{red}{Only need to write the steps and algorithmic form. I will add the algorithmic form.}}

In step 2 of the previously prescribed randomized SS scheme (Algorithm-~\ref{alg:RandSS1}) involves SVD of a matrix of size $q\times n$; hence the time complexity of the computing SVD will be $\mathcal{O}(nq^2)$, whereas House-holder transformed based QR decomposition \cite{golub2012matrix} can reduce the time complexity further. In this sub-section, we proposed a new randomized SS scheme based on the QR-decomposition. 
Suppose QR-decomposition of $\bar{\mathcal{A}}_{q \times n}$ can be factorized as
\begin{equation}
\label{Eq:qrY}
{Q_{\bar{\mathcal{A}}}}_{q \times q}\, {R_{\bar{\mathcal{A}}}}_{q \times n} \approx \bar{\mathcal{A}}
\end{equation}
Here we use partial QR-decomposition as $q$ columns of ${R_{\bar{\mathcal{A}}}}_{q \times n}$ are non-singular. 
Previously we have shown top-$q$ columns of $\mathcal{V}$ can be approximated by $\mathcal{V}_{\bar{\mathcal{A}}}$. Therefore, our aim will be to obtain $\mathcal{V}_{\bar{\mathcal{A}}}$ from the QR-factors.\\  
Let that SVD factors of $R_{\bar{\mathcal{A}}}$ can be written as
\begin{equation}
\label{Eq:svdR}
{\mathcal{U}_{R_{\bar{\mathcal{A}}}}}_{q \times q}\, {\Sigma_{R_{\bar{\mathcal{A}}}}}_{q \times q}\,{\mathcal{V}^T_{R_{\bar{\mathcal{A}}}}}_{q \times n} \approx R_{\bar{\mathcal{A}}}
\end{equation}
Combining Eq:\ref{Eq:qrY} $\&$ \ref{Eq:svdR}, we can re-write 
\begin{equation}
\label{Eq:qrsvdY}
\bar{\mathcal{A}} \approx \big(Q\,\mathcal{U}_{R_{\bar{\mathcal{A}}}}\big) \, \Sigma_{R_{\bar{\mathcal{A}}}} \, \mathcal{V}^T_{R_{\bar{\mathcal{A}}}}
\end{equation}

From Eq.\ref{Eq:svd_Y} $\&$ \ref{Eq:qrsvdY}, it is clear that 
$\mathcal{U}_{\bar{\mathcal{A}}} \approx Q\,\mathcal{U}_{R_{\bar{\mathcal{A}}}}$ ;
$\Sigma_{\bar{\mathcal{A}}} \approx \Sigma_{R_{\bar{\mathcal{A}}}}$ ;
$\mathcal{V}_{\bar{\mathcal{A}}} \approx \mathcal{V}_{R_{\bar{\mathcal{A}}}}$
Hence one can compute $\mathcal{V}_{R_{\bar{\mathcal{A}}}}$ and that can be used to approximate $\mathcal{V}_{\bar{\mathcal{A}}}$, which eventually approximates $\mathcal{V}$. These steps are written in Algorithm-\ref{alg:RandSS2}. This algorithm is not computationally efficient as it involves one QR-decomposition (step2) and another SVD factorization (step3) of a matrix of size $q \times n$. The the previously described Algorithm-\ref{alg:RandSS1} is more computationally acceptable in comparison to this algorithm. It is further possible to discard the step3 of the algorithm by slightly modifying the algorithm which described in the next section as Algorithm-\ref{alg:RandSS3}.
%%%%=====================================================

% This algorithm just alters the third step ~\ref{eqn:eqsvd} of previous step and replaces it with two steps as follows.

% \begin{equation}
% Q, R = QR(\mathcal{\bar{A}})
% \end{equation}
% and, 

% \begin{equation}
% \mathcal{U}_{R_{\bar{\mathcal{A}}}}\,\Sigma_{R_{\bar{\mathcal{A}}}} \, \mathcal{V}_{R_{\bar{\mathcal{A}}}}^T = \text{svd}(\mathcal{R_{\bar{\mathcal{A}}}})
% \end{equation}

% where $QR$ represents QR decomposition. Using QR decomposition after low dimension projection further reduces the time complexity as compared to the previous method
% (RSS1) given the fact that it has lower computation time than the SVD ({\color{blue}mention the order}). Further, the first $q$ columns of the matrix $\mathcal{V}_{R_{\bar{\mathcal{A}}}}$ are extracted and used for QR-cp factorization in the manner $\mathcal{\bar{V}}$ was used in previous algorithms. Rest of the steps remain same as the previous versions (as follows).

% \begin{equation}
%     \bar{\mathcal{V}}_{R_{\bar{\mathcal{A}}}} = \mathcal{V}_{R_{\bar{\mathcal{A}}}}[:, 0:q]
% \end{equation}

% \begin{equation}
%     Q, R, P = QR(\bar{\mathcal{V}}_{R_{\bar{\mathcal{A}}}}^T) 
% \end{equation}

% \begin{equation}
% \tilde{\mathcal{A}} = \mathcal{A}[:, P]
% \end{equation}

% \begin{equation}
% {\mathcal{A}_1} = \tilde{\mathcal{A}}[:, 0:q]
% \end{equation}

% The matrix $\mathcal{A}_1$ is the requires subset.
%%%%=====================================================
\begin{algorithm}[H]
\DontPrintSemicolon
\KwIn{Data Matrix \{$\mathcal{A}_{m\times n}$\}, Random Projection Matrix \{$\Omega_{q \times m } : \Omega_{ij} \in \mathcal{N}(0,1)\}$}
\KwOut{${\mathcal{A}_1}_{m \times q} $}
$\bar{\mathcal{A}}_{q \times m} = \Omega\mathcal{A}$ \atcp{Compression of the row-space}
$Q_{\bar{\mathcal{A}}} \; R_{\bar{\mathcal{A}}} = \text{QR}(\bar{\mathcal{A}})$ \atcp{Compute the QR decomposition of the $\bar{\mathcal{A}}$}
$\mathcal{U}_{R_{\bar{\mathcal{A}}}}\, \Sigma_{R_{\bar{\mathcal{A}}}} \, \mathcal{V}_{R_{\bar{\mathcal{A}}}}^T = \text{svd}(\mathcal{R_{\bar{\mathcal{A}}}})$  \atcp{Compute the singular value decomposition}
%$\bar{\mathcal{V}}_{R_{\bar{\mathcal{A}}}} = \mathcal{V}_{R_{\bar{\mathcal{A}}}}[:, 0:q]$ \atcp{Choose first $q$ columns of $\mathcal{V}_{R_{\bar{\mathcal{A}}}}$.}
$Q, R, P = qr(\bar{\mathcal{V}}_{R_{\bar{\mathcal{A}}}}^T)$ \atcp{Compute column-pivotal $QR$ decomposition.}
$\tilde{\mathcal{A}} = \mathcal{A}[:, P]$ \atcp{Permute the column of the data-matrix using permutation matrix $P$.}
${\mathcal{A}_1}_{m \times q} = \tilde{\mathcal{A}}[:, 0:q]$ \atcp{Choose first $q$ columns.}
\caption{\textbf{Randomized subset selection algorithm-II}}
\label{alg:RandSS2}
\end{algorithm}
%%%%====================================================
\subsection{Randomized Subset Selection 3}

Another advancement in terms of time complexity reduction is to discard the step3 of the previously described algorithm (Algorithm-\ref{alg:RandSS2}). 
From the Eq.\ref{Eq:svd_Y}, it is clear that 
\begin{equation}
\label{Eq:svdTY}
\bar{\mathcal{A}}^T \approx \mathcal{V}_{\bar{\mathcal{A}}} \, \Sigma_{\bar{\mathcal{A}}}^T \, \mathcal{U}_{\bar{\mathcal{A}}}^T 
\end{equation}
Therefore, if one can compute QR-decomposition of $\bar{\mathcal{A}}^T$ as
\begin{equation}
\label{Eq:qrTY}
Q_{\bar{\mathcal{A}}^T} \, R_{\bar{\mathcal{A}}^T} \approx \bar{\mathcal{A}}^T
\end{equation}
then from Eq.\ref{Eq:svdTY} $\&$ \ref{Eq:qrTY}, it is easy to relate 
\begin{equation}
\mathcal{V}_{\bar{\mathcal{A}}} \approx Q_{\bar{\mathcal{A}}^T}
\end{equation}

This scheme is more efficient than previously described two different randomized algorithm (Algorithm-\ref{alg:RandSS1}, \ref{alg:RandSS2}), as it involves only QR-decomposition (step2) of a matrix of size $n \times q$. 

% use QR decomposition of $\mathcal{\bar{A}}^{T}_{m \times q}$ obtained after the projection of original matrix $\mathcal{A}$ into a low dimensional space and taking its transpose as, 
% \begin{equation}
% \mathcal{\bar{A}}_{q \times m} = \Omega_{q \times m} \mathcal{A}_{m \times m}\ ,\ q<m
% \end{equation},

% \begin{equation}
% Q, R = QR(\mathcal{\bar{A}})
% \end{equation}

% Then, the resulting orthogonal $Q_{m \times q}$ matrix is used in place of $\bar{V}^T$ for $QR_{cp}$ factorization and permutation  vector and important matrix subset was obtained as in section {\color{blue}2.1.1}
% \begin{equation}
%  \bar{V} = Q_{m \times q}
% \end{equation}

% \begin{equation}
%  Q, R, P = QR_{cp}(\bar{V}^{T}_{q \times m})
% \end{equation} 

% The matrix subset is obtained by ordering columns of $\mathcal{A}$ according to the permutation vector $P$ and picking up first $q$ columns of permuted
% $\mathcal{A}$.\\
%%%%=====================================================
\begin{algorithm}[H]
\DontPrintSemicolon
\KwIn{Data Matrix \{$\mathcal{A}_{m\times n}$\}, Random Projection Matrix \{$\Omega_{q \times m } : \Omega_{ij} \in \mathcal{N}(0,1)\}$}
\KwOut{${\mathcal{A}_1}_{m \times q} $}
$\bar{\mathcal{A}}_{q \times n} = \Omega\mathcal{A}$ \atcp{Compression of the row-space}
$Q_{\bar{\mathcal{A}}^T} \; R_{\bar{\mathcal{A}}^T} = \text{QR}(\bar{\mathcal{A}}^T)$ \atcp{Compute the QR decomposition of the $\bar{\mathcal{A}}^T$}
$\bar{\mathcal{V}} = Q_{\bar{\mathcal{A}}^T}^T$\\
$Q, R, P = qr(\bar{\mathcal{V}})$ \atcp{Compute column-pivotal $QR$ decomposition.}
$\tilde{\mathcal{A}} = \mathcal{A}[:, P]$ \atcp{Permute the column of the data-matrix using permutation matrix $P$.}
${\mathcal{A}_1}_{m \times q} = \tilde{\mathcal{A}}[:, 0:q]$ \atcp{Choose first $q$ columns.}
\caption{\textbf{Randomized subset selection algorithm-III}}
\label{alg:RandSS3}
\end{algorithm}
%%%%====================================================
\subsection{Randomized Subset Selection 4}

This section describes the fourth randomized version of the algorithm for subset selection. The previous three versions (Algorithm-\ref{alg:RandSS1}, \ref{alg:RandSS2}, \ref{alg:RandSS3}) have used SVD or QR factorization of a matrix of size $q \times n$. But, if $n$ is also very large then time complexity will not be reduced to the desired level. Therefore, we are interested in transforming the data matrix to a much smaller size and retrieval of the first $q$-right singular vectors approximately. The prescribed algorithm in this section is based on constructing a matrix $Y$ of size $q \times q$ and approximates right singular vectors correctly. The transformation from $\mathcal{A}_{m \times n} \rightarrow \bar{\mathcal{A}}_{q \times n} \rightarrow {Y}_{q \times q}$ is logical and can be explained through simple algebraic relations. The first transformation ($\mathcal{A}_{m \times n} \rightarrow \bar{\mathcal{A}}_{q \times n}$), as described in the previous section, showed that the approximation of right singular vectors of $\mathcal{A}$ is possible by the right singular vectors of $\bar{\mathcal{A}}$. In this sub-section, we discuss the second transformation ($\bar{\mathcal{A}}_{q \times n} \rightarrow {Y}_{q \times q}$) and the retrieval of right singular vectors of $\bar{\mathcal{A}}$ from $Y$. 
From Eq.\ref{Eq:svd_Y}, it is trivial to obtain,
\begin{equation}
\label{Eq:AAT}
Y_{q \times q} = \bar{\mathcal{A}}\, \bar{\mathcal{A}}^T = \mathcal{U}_{\bar{\mathcal{A}}}\,\Sigma^2_{\bar{\mathcal{A}}}\, \mathcal{U}^T_{\bar{\mathcal{A}}}
\end{equation}
The above Eq.\ref{Eq:AAT} shows that $Y$ preserves the information about the left-singular vectors $\mathcal{U}_{\bar{\mathcal{A}}}$ and it reflects a eigenvalue relation of $Y$. 
\begin{equation}
\label{Eq:eigAAT}
Y \, \mathcal{U}_{\bar{\mathcal{A}}} = \Big(\bar{\mathcal{A}}\, \bar{\mathcal{A}}^T\Big)\mathcal{U}_{\bar{\mathcal{A}}} = \mathcal{U}_{\bar{\mathcal{A}}}\,\Sigma^2_{\bar{\mathcal{A}}}
\end{equation}
Therefore, one can compute the eigenvalue decomposition of $Y$ and obtain the approximated left singular vectors, singular values of $\bar{\mathcal{A}}$. 
Suppose the eigenvalue decomposition of $Y$ is 
\begin{equation}
\label{Eq:eigY}
Y X = \Lambda X
\end{equation}
Therefore comparing Eq.\ref{Eq:eigAAT} $\&$ \ref{Eq:eigY}, we can get
\begin{equation*}
\mathcal{U}_{\bar{\mathcal{A}}} \approx X
\end{equation*}
\begin{equation*}
\Sigma_{\bar{\mathcal{A}}} \approx \sqrt{\Lambda}
\end{equation*}

The right singular vector of $\bar{\mathcal{A}}$ can be defined as
\begin{equation}
\mathcal{V}_{\bar{\mathcal{A}}} = \bar{\mathcal{A}}^T \, \mathcal{U}_{\bar{\mathcal{A}}}\, \Sigma_{\bar{\mathcal{A}}}^{-1} \approx \bar{\mathcal{A}}^T \, X\, \Lambda^{-1}
\end{equation}

%%%%=====================================================
% After the random projection of columns of $\mathcal{A}$, a square matrix $Y$ is obtained by its inner product as follows,

% \begin{equation}
% Y_{q \times q} = \bar{\mathcal{A}}\bar{\mathcal{A}}^T
% \end{equation}

% Further eigenvalue decomposition of the $Y$ matrix is performed, followed by construction of $\tilde{\mathcal{V}}$ as follows:
% \begin{equation}
% \Lambda , X = \text{eig}(Y)
% \end{equation}

% \begin{equation}
% \tilde{\mathcal{V}} = \sqrt{\Lambda}\, X \, \bar{\mathcal{A}}
% \end{equation}

% Next, the first $q$ columns of $\tilde{\mathcal{V}}$ are used to obtain \bar{\mathcal{V}}, on which QR-cp factorization is done to obtain the permutation vector $P$.
% \begin{equation}
% \bar{\mathcal{V}} = \tilde{\mathcal{V}}[:, 0:q] 
% \end{equation}

% \begin{equation}
% Q, R, P = qr(\bar{\mathcal{V}}) 
% \end{equation}
%%%%=====================================================
\begin{algorithm}[H]
\DontPrintSemicolon
\KwIn{Data Matrix \{$\mathcal{A}_{m\times n}$\}, Random Projection Matrix \{$\Omega_{q \times m } : \Omega_{ij} \in \mathcal{N}(0,1)\}$}
\KwOut{${\mathcal{A}_1}_{m \times q} $}
$\bar{\mathcal{A}}_{q \times n} = \Omega\mathcal{A}$ \atcp{Compression of the row-space}
$Y_{q \times q} = \bar{\mathcal{A}}\bar{\mathcal{A}}^T$\\
$\Lambda , X = \text{eig}(Y)$\atcp{Eigenvalue decomposition of $Y$}
$\tilde{\mathcal{V}} = \bar{\mathcal{A}}^T \, X\, \Lambda^{-1}$ \\
%$\bar{\mathcal{V}} = \tilde{\mathcal{V}}[:, 0:q]$ \atcp{Choose first $q$ columns of $\tilde{\mathcal{V}}$.}
$Q, R, P = qr(\bar{\mathcal{V}}^T)$ \atcp{Compute column-pivotal $QR$ decomposition.}
$\tilde{\mathcal{A}} = \mathcal{A}[:, P]$ \atcp{Permute the column of the data-matrix using permutation matrix $P$.}
${\mathcal{A}_1}_{m \times q} = \tilde{\mathcal{A}}[:, 0:q]$ \atcp{Choose first $q$ columns.}
\caption{\textbf{Randomized subset selection algorithm-IV}}
\label{alg:RandSS4}
\end{algorithm}
%%%%=====================================================
\vspace{0.25cm}

%%%%=====================================================
\begin{figure*}[htbp]
\centering
\begin{tabular}{cc}
 \centering
 \includegraphics[height=5cm]{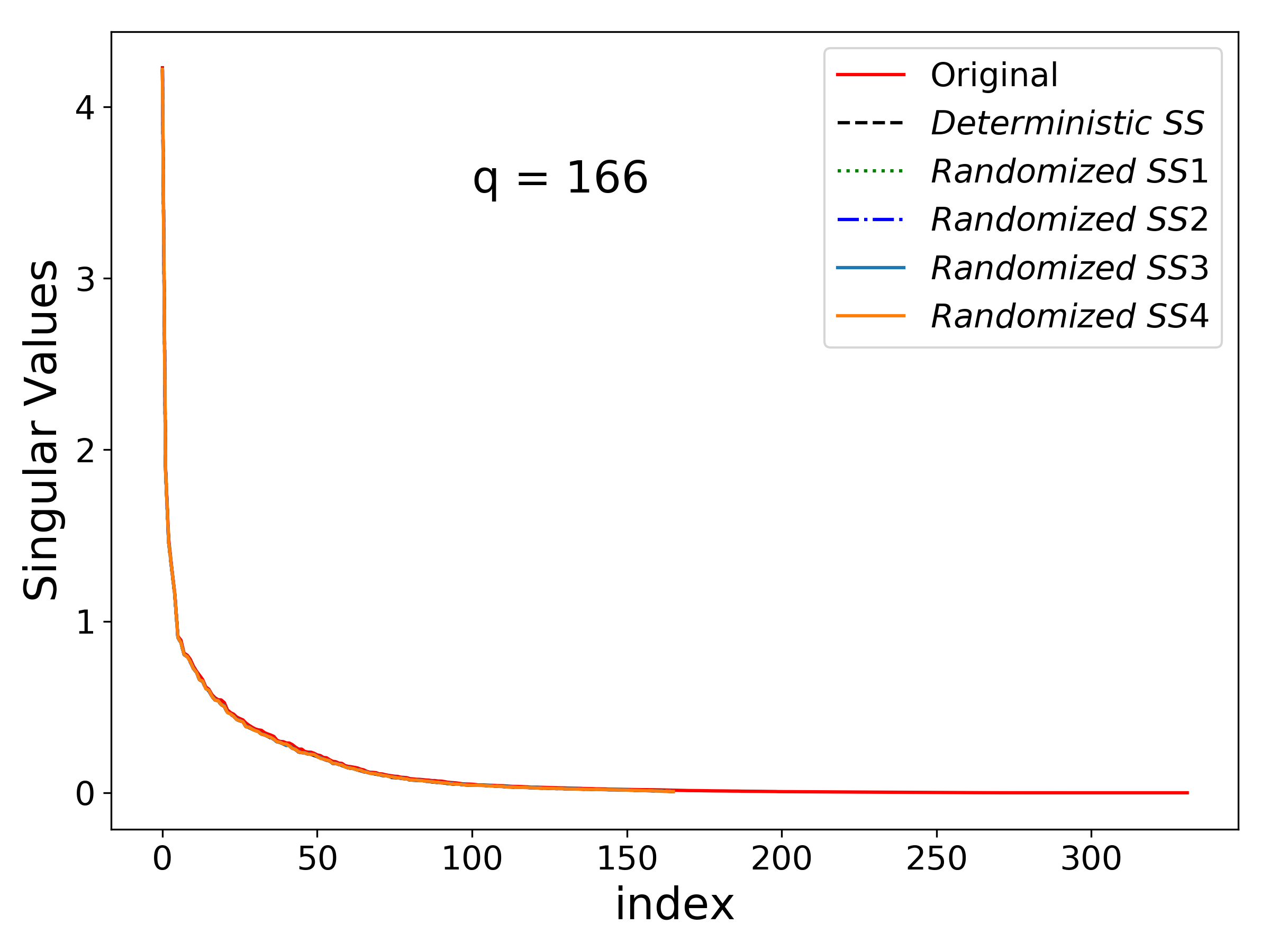}&
 \includegraphics[height=5cm]{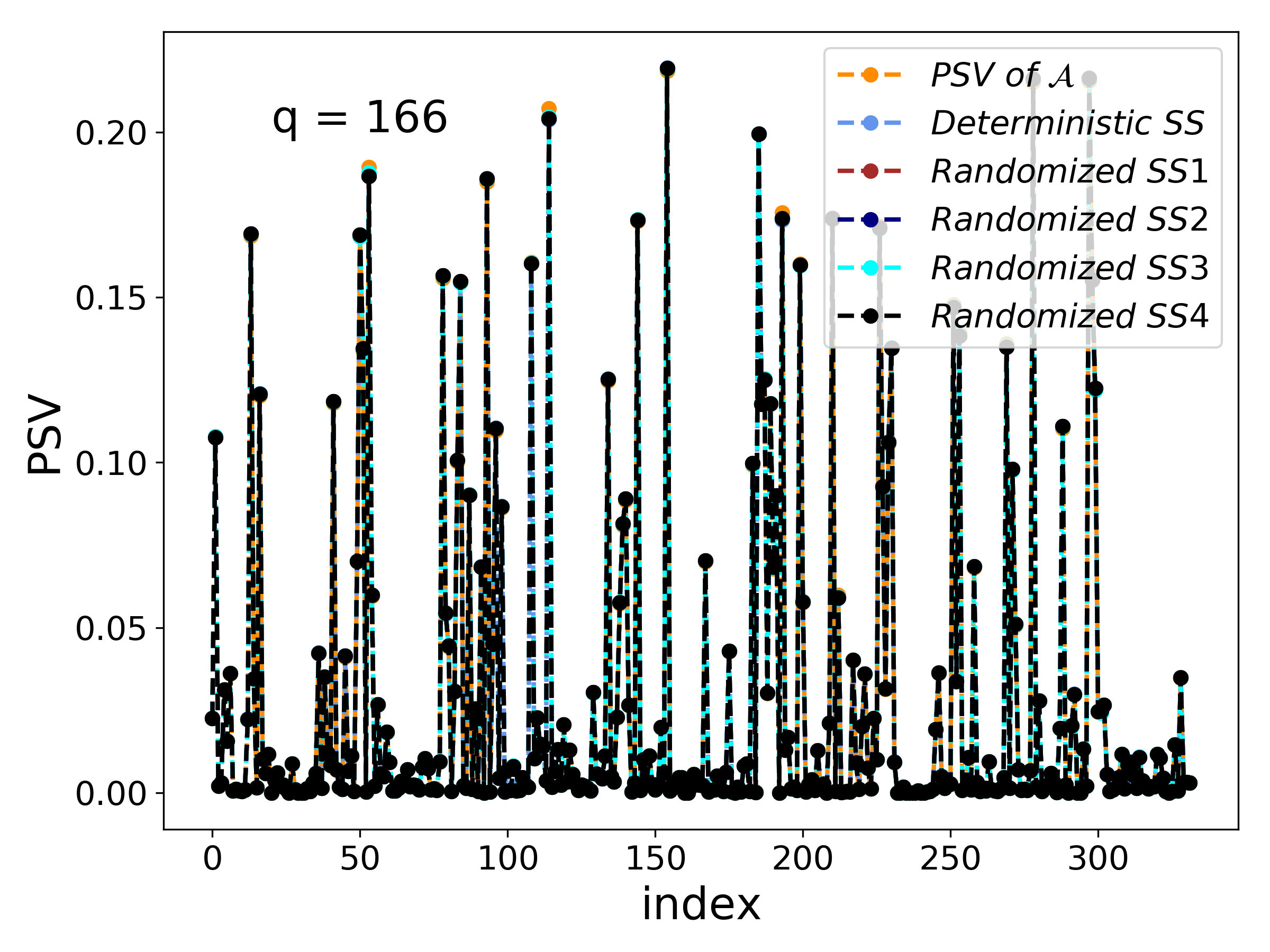}\\
 (a)&(b)
 \end{tabular}
 \caption{(a) The figure shows singular value spectra of the US Air network (with 322 nodes and 2126 edges) and the subsets with $q = 166$ extracted through various algorithms proposed in the paper. (b) The figure shows the plot of Principal singular Vectors of US Air network adjacency matrix and that of subsets extracted through various algorithms. The excellent overlap between the PSVs of main network and subsets show that subset retains the spectral information of the main network.}
\label{fig:cs1}
 \end{figure*}
%%%%%%%%%%%%%%%%%%%%%%%%%%%%%%%%%%%%%%%%%%%%%%%%%%%%%%%%%%%%%%%%%%%%
\subsection{Computational complexity analysis}
%{\color{red}{I will write the floating point comparison of the methods. Also we need to make a table based on the run-time comparison of these decomposition methods. I will make a futuristic plot to give a overview of TC for large data matrix.}}

To understand the computational complexity of proposed Algorithm \ref{alg:RandSS1}-\ref{alg:RandSS4}  in terms of the number of floating points operations, we evaluate the complexity of required steps. We mentioned earlier that the computation of column pivoting QR-decomposition of the truncated right singular matrix is the essential step for the all randomized and deterministic SS method. Therefore, in the computational complexity analysis, we excluded that cost. 
All randomized algorithms involves a multiplication of the data matrix with a random projection matrix, therefore it requires $\mathcal{O}(mnq)$. 
Apart from that, the Algorithm \ref{alg:RandSS1} involves a SVD factorization of a matrix of size $q \times n$ which cost $\mathcal{O}(nq^2)$. Similarly, the Algorithm \ref{alg:RandSS2}, \ref{alg:RandSS3} involves a QR-factorization of a matrix of $q\times n$ which cost $\mathcal{O}(2nq^2 - \frac{2}{3}q^3)$, considering the House-Holder transformation. Whereas, the Algorithm \ref{alg:RandSS4} involves a eigenvalue decomposition in step3, therefore the time complexity is $\mathcal{O}(q^3)$, which is very cheap for a low-rank structure in comparison to the other schemes.

\begin{table*}
\begin{center}
\label{Tab:TimeCom}
\begin{tabular}{ | m{3cm} | m{3cm}| m{3cm} | } 
\hline
Algorithm & Matrix multiplication & SVD/QR \\ 
\hline
Algorithm \ref{alg:DetSS}& - & $\mathcal{O}(mn^2)$ \\ 
\hline
Algorithm \ref{alg:RandSS1} & $\mathcal{O}(mnq)$ &$\mathcal{O}(nq^2)$  \\ 
\hline
Algorithm \ref{alg:RandSS2} & $\mathcal{O}(mnq)$ &$\mathcal{O}(2nq^2 - \frac{2}{3}q^3)  + \mathcal{O}(nq^2) $  \\
\hline
Algorithm \ref{alg:RandSS3} & $\mathcal{O}(mnq)$ &$\mathcal{O}(2nq^2 - \frac{2}{3}q^3)$  \\
\hline
Algorithm \ref{alg:RandSS4} & $\mathcal{O}(mnq) + \mathcal{O}(nq^2)$ &$\mathcal{O}(q^3)$  \\
\hline
\end{tabular}
\end{center}
\caption{This table shows the theoretical time complexity involved in all the algorithms proposed in the paper.}
\end{table*}

\begin{table*}
\begin{center}
\label{Tab:TimeCom}
\begin{tabular}{ | m{2cm} | m{2cm}| m{2cm} | m{2cm} | m{2cm} | m{2cm} | m{2cm} | } 
\hline
Algorithm & \textbf{BA} & \textbf{Drosophila} & \textbf{ER} & \textbf{Friendship} & \textbf{Power Grid} & \textbf{US Air} \\ 
\hline
Algorithm \ref{alg:DetSS}& 1.014 & 4.895 & 1.148 & 5.165 &87.408 & 0.060\\ 
\hline
Algorithm \ref{alg:RandSS1} & 0.371 & 1.761 & 0.380 & 1.978 & 32.546& 0.034\\ 
\hline
Algorithm \ref{alg:RandSS2} & 0.392  & 2.097 & 0.434 & 2.306 & 37.424& 0.036\\
\hline
Algorithm \ref{alg:RandSS3} & 1.109 & 4.116 & 1.185 & 4.364 & 56.132& 0.057\\
\hline
Algorithm \ref{alg:RandSS4} & 0.281 & 1.268 & 0.252 & 1.441 & 55.639 & 0.044 \\
\hline
\end{tabular}
\end{center}
\caption{The table shows the numerical time complexity involved in subset selection procedure for all the algorithms proposed in the paper for six networks.}
\end{table*}

The TABLE I. shows the detail of the complex complexity of all the described algorithms. The Algorithm \ref{alg:RandSS4} is more efficient in comparison to the other proposed randomized SS scheme in terms of practical applicability of the scheme. This scheme is highly parallelizable and suited for the modern computer architecture. This algorithm involves only one eigenvalue decomposition (step3) of a matrix of dimension $q \times q$. Due to the low-rank structure, the value of $q$ will be very small for a large data matrix, therefore this eigenvalue operation can be easily done on a single machine, whereas the other randomized algorithms involve a SVD/ QR-factorization of $q\times n$ sized matrix, therefore for a large $n$, single machine can not perform this operation. 

If we consider the total time complexity of each of prescribed algorithm then the Algorithm\ref{alg:RandSS3} is more cost efficient as Algorithm \ref{alg:RandSS4} involves a huge matrix multiplication which has a cost of $\mathcal{O}(nq^2)$. But all the prescribed algorithms are useful for a large data matrix, therefore we can't store the data matrix into one specific machine, hence we have to store the data in a distributed architecture. In distributed architecture, the matrix multiplication cost will be reduced drastically. Hence, cost of the algebraic operations should not be included to compare the computationally efficiency of the methods. It would be better to provide a computational complexity analysis of these algorithms in a distributed system which is beyond the scope of this paper. Therefore the comparison based on the third column of the TABLE I makes more sense as performing SVD or QR on a large dimensional data matrix is the bottleneck. Based on that comparison, all the randomized algorithm out perform the complex complexity obtain by the deterministic SS.    \\

In the TABLE II, we show the time required to obtain the subset using all the deterministic and randomized SS methods for six networks. The values of time units are averaged over independent 20 trials. If $q<<n$, it is expected that the time required will be much less that the value in the TABLE II, which is for $q \approx n/2$. \\

%%%%=====================================================
%%%%=====================================================
\begin{figure*}[htbp]
\centering
\includegraphics[height=10cm, trim={10 60 0 60},clip]{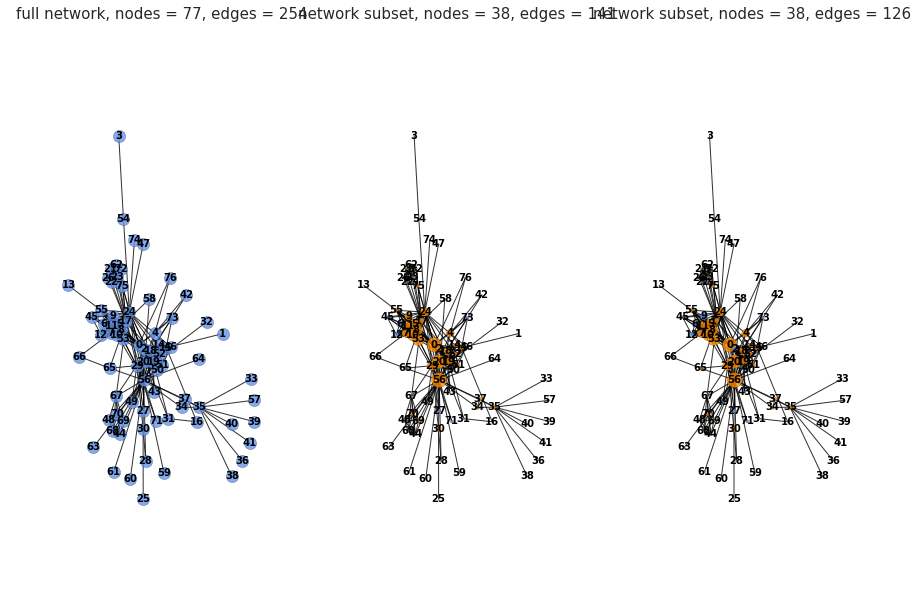}
\caption{The figure in leftmost panel shows the Les Miserables network \cite{kunegis2013konect} with 77 nodes and 254 edges. The middle figure shows the same network with subset network (38 nodes in orange color, 141 edges) extracted using deterministic SS process embedded in the main network. The third network shows the subset network (38 nodes in orange color, 126 edges)extracted using randomized SS1 process embedded in the main network. Please note that the base network nodes are scaled according to their eigenvector centralities in second and third panels. The orange  color nodes being large in size implies that subset captures the most central nodes in the network.}
\label{fig:cs2}
\end{figure*}
%%%%=====================================================
\begin{figure*}[htbp]
\centering
\includegraphics[height=10cm, trim={10 60 0 60},clip]{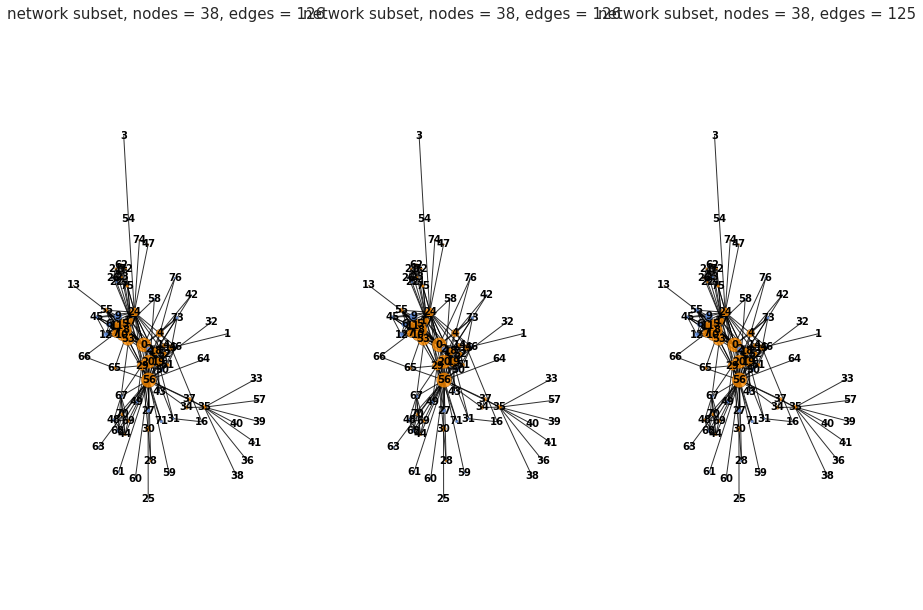}
\caption{The figure in leftmost panel shows embedded subset network in the the Les Miserables network \cite{kunegis2013konect} with 77 nodes and 254 edges. The subset network is extracted using Randomized SS2 has 38 nodes (shown in orange), 126 edges. The middle figure shows the same network with subset network (38 nodes in orange color, 126 edges) extracted using Randomized SS3 process embedded in the main network. The third network shows the subset network (38 nodes in orange color, 125 edges)extracted using randomized SS4 process embedded in the main network. Please note that the base network nodes are scaled according to their eigenvector centralities in second and third panels. The orange  color nodes being large in size implies that subset captures the most central nodes in the network.}
\label{fig:cs3}
\end{figure*}

\section{Application of SS selection procedure in complex networks}
%{\color{red}{Summarize the outcome of our previous work and why it is important.}}
For a detailed account of SS procedure application on complex networks, please refer \cite{tripathi2019subset}. However, for the sake of completeness, we summarise it here. The essential condition or constraint to obtaining size reduced representation of complex networks is to retain its eigenvalue spectra and to quantify any loss in matrix energy in terms of Frobenius norms difference of the $\mathcal{A}$ and the $\mathcal{A}_1$. On a slightly different note, the constraint conditions, i.e. preservation of full matrix energy in the subset and overlap of PSVs, may not be imposed on the SS procedure at all. For this case, one computes subset for an arbitrary value of $q$, and the subset has $q$ most linearly independent columns. In \cite{tripathi2019subset}, we showed that even when the subset size (number of columns = $q$) was chosen to be half the size of $\mathcal{A}$, the PSV of $\mathcal{A}_1$ has more than 99$\%$ overlap with PSV of $\mathcal{A}$ also the loss of matrix energy was very minimal.\\

The amount of overlap between PSVs was quantified using Cosine Similarity (CS), which is a measure of relative orientations of the two vectors (refer eq.~\ref{eq:cs}, where $\mathcal{A}^{1}$ and $\mathcal{A}_{1}^{1}$ are PSVs of main adjacency matrix and the subset). Bounded between [0, 1], CS is maximum when the two vectors are oriented along the same direction and minimum when they are perpendicular to each other. In our case, the CS is a measure of the extent of information retrieval from the main matrix into the subset.  \\
 
 \begin{equation}
 CS = cos(\theta) = \frac{\mathcal{A}^{1}.\mathcal{A}_{1}^{1}}{\parallel \mathcal{A}^{1} \parallel \parallel \mathcal{A}_{1}^{1} \parallel}
 \label{eq:cs}
\end{equation}

Obtaining the subset network is not very straightforward as the selected subset is generally a non-square matrix. To this end, the subset selection procedure was extended, and the rows of the selected subset were reordered using same $P$(Refer eq.~\ref{eqn:eqB}, $\mathcal{B}$ has top $q$ rows, and $\mathcal{B}_r$ has remaining ones). This is intuitively justified owing to the symmetric nature of the adjacency matrix. Hence the subset network has $q$ nodes with connections defined by subset matrix $\mathcal{B}$.

\begin{equation}
   [\mathcal{B}^T, \mathcal{B}_r^T] \equiv P\mathcal{A}_1
   \label{eqn:eqB}
\end{equation}

We find that subsets extracted through all the randomized versions of the deterministic SS procedure have excellent CS of 
Principal Singular Vector (PSV) with that of the original network adjacency matrix (see FIG.~\ref{fig:cs1}. This has two important implications. Firstly, 
the conservation of PSV in subsets implies that one can infer the influence of nodes from the subset PSV itself. Also, as the PSV entries are indicative 
of node influence (components of PSV being the eigenvalue centralities of the nodes), one can infer a node's influence in spreading processes on the 
network using the subset PSV's corresponding component. This is a good result, as doing the SVD of a large network to find PSV can be computationally 
expensive. Secondly, the subset network identifies the most important network structure in terms of its nodes and edges, such that selected network has
enhanced information flow governing properties. We have shown the application of SS using all the deterministic and randomized SS on US Air 
network \cite{kunegis2013konect} in FIG.~\ref{fig:cs2} and FIG.~\ref{fig:cs3}. One can see that all the deterministic and randomized algorithms can 
efficiently detect almost all high eigenvector centrality nodes and hence we can conclude that the subset nodes contribute maximally to the inverse 
participation ratio of the networks. Please refer to \cite{tripathi2019subset} for details. Also, we find that the loss in matrix norm of the subsets 
with $q \approx n/2$ is very minimal for all the network examples we took (refer TABLE III).

%%%%=====================================================

%%%%=====================================================
%\section{Application to the complex network}

\begin{table*}[h]
\label{tab:tab1}
\begin{center}
    \begin{tabular}{| p{2.7cm} || p{2.3cm} | p{2cm} | p{1cm} | p{2cm} | p{1cm} |p{1cm} |p{1cm} |p{1cm} |p{1cm} | p{1cm} |}
    \hline
    Type & Networks &  (V, E) of network  & q & (V, E) of SS network & loss1 &loss2 & loss3 & loss4 & loss5 & $CS$\\
    \hline
    \multirow{3}{*}{Weighed, real}& US Air & (332, 2126) & 166 &                                       (890,1635) & 0.011 & 0.010 & 0.010& 0.010 & 0.010 & $>$0.99\\
                       &Les Miserables & (77, 254) & 38 & 
                       (38, 141) & 0.017 & 0.028 &0.028 & 0.028 & 0.025 & $>$0.99\\
                       &Train Bombing & (64, 243) & 32 & 
                       (32, 123) & 0.100& 0.100& 0.100& 0.100& 0.105 & $>$0.99\\
                       \hline
    \multirow{6}{*}{Unweighted, real}&Karate & (34, 78) & 20 & (20, 47)& 
                                     0.156& 0.164& 0.164& 0.164& 0.152 & $>$0.94 \\
                        &Cat Brain & (65, 730) & 32 & (32, 247) &
                        0.236& 0.232& 0.232& 0.232& 0.219 & $>$0.99 \\
                        &Drosophila & (1781, 9016) & 890 & (890, 7026) & 
                        0.057& 0.058& 0.058& 0.058& 0.056 & $>$0.99 \\
                        &Power Grid & (4941, 6594) &  2470 & (2470, 2863) &
                        0.166& 0.161& 0.161& 0.161& 0.160 & $>$0.985 \\
                        &Jazz Musicians& (198, 2742) & 96 & (96, 947) &
                        0.232& 0.214& 0.214& 0.214& 0.220 & $>$0.98 \\
                        &Friendship & (1858, 12534) & 929 & (929, 7618) & 
                        0.114& 0.113& 0.113& 0.113& 0.109 & $>$0.99 \\
                        \hline
    \multirow{3}{*}{Unweighted, model}&Barabasi Albert & (1000, 2991) & 500 & (500,                                   1418) & 0.150& 0.151& 0.151& 0.151& 0.151 & $>$0.99  \\
                        &Erdos Renyi & (1000, 7558) & 500 & (500, 2630) &
                        0.230& 0.234& 0.234& 0.234& 0.233 & $>$0.98 \\
                        &Power Law & (1000, 1360) & 500 & (500, 947) &
                        0.082& 0.087& 0.087& 0.087& 0.082 & $>$0.88 \\

    % N6 &  & & & & & &\\
    % N7 &  & & & & & &\\
    % N8 &  & & & & & &\\
    % N9 &  & & & & & &\\
    % N10 &  & & & & & &\\
    \hline
    \end{tabular}
\end{center}
\caption{A table of SS results on model networks and weighted and unweighted real networks examples. The real networks were downloaded from KONECT \cite{kunegis2013konect} and model network types were generated using python module Networkx \cite{hagberg2008exploring}. (V, E) represents the vertices and the edges in the networks. The columns loss1, loss2, loss3, loss4 and loss5 represent the Frobenius norm differences of main networks and the corresponding subsets obtained through Deterministic SS, RSS1, RSS2, RSS3 aand RSS4 respectively. CS represents cosine similarity between PSVs of main network adjacency and all the subsets. It is found to be greater than the value in the column for all the subsets.}
\end{table*}

%%%%=====================================================
\section{Conclusion and discussion}
The present manuscript presents a class of randomized subset selection procedures on complex networks data. The main highlight of this work is the use of Random Projection scheme in the process of extract top $q$ most linearly independent columns from a data matrix. The RP method thrives on rank deficiency of input data matrix. Apart from reducing the time complexity incurred due to the performing of SVD of full data matrix required for deterministic SS procedure, RP can preserve the maximum information from the data matrix. We have verified the applicability of the proposed methods on complex network datasets. The complex networks, for example, the Internet, World Wide Web and traffic network,  can be huge as well as dynamically evolving. The proposed methods owing to there reduced time complexity can unburden the computing devices a lot on such datasets. Finding the spectra and eigenvectors of complex networks is of paramount importance to infer its topological and functional properties. Also, finding the most critical nodes and links or the most functional network structure is currently one of the most researched topics in complex networks. We showed that using the SS procedure, the important network structure can be extracted which captures the most influential network nodes. Using randomized version of SS, this process can be fastened by many folds. We have taken $q \approx N/2$, and showed all the results with this $q$ itself. However, $q$ can further be reduced depending on sparsity of data (small Network density) and time complexity can further be reduced. The determination of appropriate $q$ is altogether a different problem and serves as a prelude to our work. Although we have applied the randomized SS procedure to complex network data, these procedures can very well be extended to the general class of large datasets and real-time analysis of time evolving data.

%%%%=====================================================
\bibliography{References}% Produces the bibliography via BibTeX.

\end{document}